\documentclass[%
reprint,
amsmath,amssymb,
aps,
]{revtex4-2}

\usepackage{graphicx}
\usepackage{dcolumn}
\usepackage{bm}

\usepackage{makecell}
\usepackage{caption}
\usepackage{subcaption}
\usepackage[colorlinks,linkcolor=blue,citecolor=blue,urlcolor=blue,bookmarks=false,hypertexnames=true,linktocpage=true]{hyperref}
\usepackage{xcolor}

\newcommand{\figref}[1]{Fig.~\ref{fig:#1}}

\newcommand{\refme}[1][]
{\textbf{\textcolor{red}{[r\ifthenelse{\isempty{#1}}{}{:#1}]\ }}}

\begin{document}	
	\title{Coherent feedback for quantum expander in gravitational wave observatories}
	
	\author{Niels B\"ottner}
 	\author{Joe Bentley}
	\author{Roman Schnabel}
 	\author{Mikhail Korobko}
	\email{mkorobko@physnet.uni-hamburg.de}
	\affiliation{Institut f\"ur Quantenphysik und Zentrum für Optische Quantentechnologien , Universit\"at Hamburg, Luruper Chaussee 149, 22761 Hamburg, Germany}
	
	\date{\today}
	
	\begin{abstract}
		The observation of gravitational waves from binary neutron star mergers offers insights into properties of extreme nuclear matter. However, their high-frequency signals in the kHz range are often masked by quantum noise of the laser light used. Here, we propose the ``quantum expander with coherent feedback'', a new detector design that features an additional optical cavity in the detector output and an internal squeeze operation. This approach allows to boost the sensitivity at high frequencies, at the same time providing a compact and tunable design for signal extraction. It allows to tailor the sensitivity of the detector to the specific signal frequency range. We demonstrate that our design allows to improve the sensitivity of the high-frequency detector concept NEMO (neutron star extreme matter observatory), increasing the detection rates by around 15\%. Our approach promises new level of flexibility in designing the detectors aiming at high-frequency signals.
	\end{abstract}
	
	\maketitle

	\section{\label{sec:Introduction}Introduction}
    
    Since the first detection of gravitational waves (GWs) from a binary black hole merger in 2015~\cite{Abbott2016}, the field of gravitational-wave observation has entered a new phase where events are detected every few days~\cite{TheLIGOScientificCollaboration2017a, TheLIGOScientificCollaboration2019,LIGO2022O31st,LIGO2023O32nd}.
    The anticipated new generation of GW detectors, such as Einstein Telescope~\cite{Branchesi2023} and Cosmic Explorer~\cite{Evans2021}, will be so sensitive, that they are predicted to sense multiple events per minute. 
    These detectors, like the current generation~\cite{AdvancedLIGO15,Acernese2020,Akutsu2019KAGRA}, focus on detecting signals at frequencies below $1\,\text{kHz}$. At frequencies above, the detectors are not sensitive enough to observe the post-merger effects that produce signals in the several kHz frequency range.
    This information, however, is crucial for understanding the fundamental properties of the ultra-dense matter in neutron stars~\cite{Read2009, Bauswein2012, Harry2018} or possible deviations from general relativity~\cite{Baiotti2017,Conklin2018}.

    The sensitivity of detectors at high frequencies is limited by the optical bandwidth of cavities used to enhance the light and signal powers.
    Modern detectors feature arm cavities, as well as additional cavities to enhance the light power (power recycling cavity, PRC) and increase the detection bandwidth (signal extraction cavity, SEC).
    The signal is enhanced by cavities within the detector bandwidth up to $\sim500$\,Hz, but suppressed at high frequencies above $\sim1\,\text{kHz}$ (HF).
    In order to further increase the sensitivity, current detectors employ frequency-dependent quantum squeezed light~\cite{ganapathy2023broadband} suppressing quantum noise in a broad frequency band~\cite{Schnabel2017}.
    This approach however does not increase the detection bandwidth~\cite{Kimble2000}.
    One way to overcome this limitation is to design a detector specifically targeting HF range by creating an additional optical resonance at kHz frequencies.
    This can be achieved by three different ways: by detuning the detector from its resonance~\cite{Ganapathy2021,Gardner2023holevo}, by increasing the length of the SEC, as planned for the proposed neutron star extreme matter observatory (NEMO)~\cite{Ackley2020}, and by adding a second long SEC in a detuned detector~\cite{Thuering2007, Thuring2009, Graf2013}.
    With the HF resonance, the energy exchange between the SEC and arm cavities occurs at the characteristic frequency of a few kHz (called the sloshing frequency~\cite{Chen2013}), and the signal at these frequencies is resonantly enhanced.
    Long SEC promises high sensitivity at HF, but also comes at a price of increased complexity and costs of the experiment. 
    At the same time, it restricts the peak operation range to 1-3\,kHz signals, limiting the sensitivity towards lower frequency signals.

    Several alternative approaches appeared in the recent years, which enhance the detection bandwidth by employing active elements directly inside the detector.
    This can be achieved in a variety of ways~\cite{Wicht1997, Wicht2000, Wise2004, Miao2015, page2021gravitational, wang2022boosting}, and we focus on the internal squeezing approach, where quantum correlated light is produced directly inside the detector, either in a degenerate~\cite{Rehbein2005, Peano2015, Korobko2017, Korobko2023, Korobko2023fundamental} or non-degenerate way~\cite{Gardner2022}.
    In order to enhance the detection bandwidth, internal squeezing should be combined with the use of HF sloshing resonance, creating a quantum expander (QE) configuration~\cite{korobko2019}.
    It can be effectively used both with short and long SEC~\cite{adya2020quantum}, but in the latter case suffers from the challenges of long SEC.

\begin{figure}
		\includegraphics[width=3.375in]{"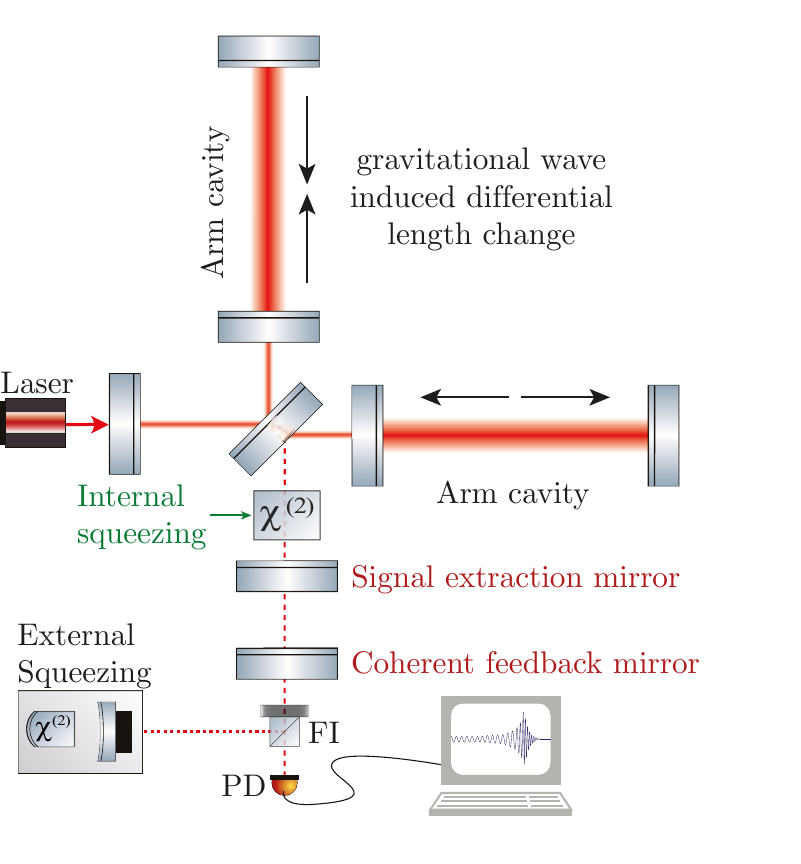"}
		\caption{\label{fig:TEQE-setup} Conceptual setup of the Quantum Expander with coherent feedback realized by the additional coherent feedback mirror. Internal squeezing is generated by pumping the nonlinear $\chi^{(2)}$ crystal with frequency doubled light. External squeezing is injected into the detector through a Faraday isolator (FI). The differential change in length of the arm cavities due to a gravitational wave, going into the page, is measured on the signal port with a photodiode (PD).}
	\end{figure}

    In this work, we increase the flexibility of the long SEC when used with QE using coherent feedback approach~\cite{yan2011coherent,zhou2015quantum,pan2018experimental,Yokoyama2020}.
    Coherent feedback allows to modify quantum state of the detector without using any additional measurements.
    In our proposal, is realized via an additional short optical cavity at the output of the detector.
    The combined Quantum Expander with coherent feedback (QECF), shown in~\figref{TEQE-setup} keeps the expanded detection bandwidth feature of the QE but provides better flexibility in choosing the frequency at which the sensitivity is maximally enhanced.
    Importantly, it significantly decreases the required SEC length by optimizing the resonance condition in the two cavities.
    We demonstrate how our approach allows to tailor the sensitivity of the detector to a specific signal frequency range.
    We compare our approach to the NEMO configuration and show the potential amplitude sensitivity gain of $\sim15\%$ at the optical resonance, and the increase of detection band towards lower frequencies.
    We perform a statistical analysis of the expected detection rates for neutron star mergers and find that the overall improved sensitivity in the kilohertz-band leads to a median gain in detection rates of $\sim15\%$ compared to NEMO.
    Improved low-frequency sensitivity allows QECF to contribute to the detection of pre-merger signals in a global detector network of LIGO~\cite{AdvancedLIGO15}, Virgo~\cite{Acernese2020}, \mbox{KAGRA}~\cite{Akutsu2019KAGRA} and other planned detectors.

 
	\section{\label{sec:TEQE}Quantum Expander with coherent feedback}

\begin{figure}
		\includegraphics[width=3.375in]{"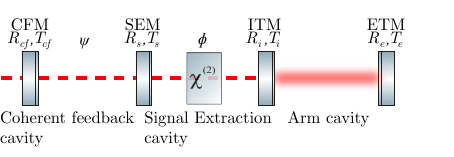"}
		\caption{\label{fig:TEQE-setup-theory_wo_fields}Simplified theoretical setup for the Quantum Expander with coherent feedback. $R_{cf,s,i,e}$ and $T_{cf,s,i,e}$ are amplitude reflectivities and transmissivities of the coherent feedback (CFM), signal extraction (SEM), input (ITM) and end mirrors (ETM) respectively. $\tau_{\text{arm},\text{SE},cf}=L_{\text{arm},\text{SE},cf}/c$ is the single trip time in the arm cavity of length $L_{\text{arm}}$, SEC of length $L_{\text{SE}}$ and CFC of length $L_{cf}$. $\psi$ is the single trip tuning of the coherent feedback cavity and $\phi$ is the single trip tuning of the signal extraction cavity.}
	\end{figure}

In this section we analytically study the response of a GW detector to quantum noise. The detector tuned close to the dark port opertation, when no carrier light exits the signal port, reflects all quantum noise entering this port directly back. In this case, the response of the detector to quantum noise is the same as of an equivalent system of coupled cavities, which we study here~\cite{Buonanno2003}. Arm cavity is formed by the input and end test masses, signal extraction cavity (SEC) is formed by the input test mass, and coherent feedback cavity (CFC) is formed by the signal extraction and coherent feedback mirrors, see \figref{TEQE-setup-theory_wo_fields}. 
All cavities are tuned in a specific way, either on resonance or off resonance.
The signal is generated as the phase modulation at the end mirror of the arm cavity, resonates in the cavities (depending on their tuning) and is then detected together with quantum noise at the output of the detector.
The sensitivity of the detector is then characterised as the noise spectral density normalized by the signal power transfer function, representing the noise-to-signal ratio at every Fourier frequency.
The QECF allows sensitivity enhancement at HF, where the detector is predominantly limited by photon shot noise.
Here, we present the main analytical result without considering the effects of quantum radiation pressure noise (QRPN) and optical losses.
Full treatment including these effects, which was used to produce the sensitivity curves in this paper, follows the derivation in~\cite{Korobko2020} and can be found in the Supplementary Material.
 
    We start by writing the input-output relations for the fields inside the cavity, and focus on the phase quadrature containing the signal $\hat{b}$~\cite{Caves1985}, 
    which is split into quantum noise and signal contributions:
	\begin{equation}
		\hat{b}=\mathcal{R}_{a}(\Omega)\hat{a}+\mathcal{T}_{x}(\Omega)x,
		\label{eq:bsol}
	\end{equation}
 where $\mathcal{R}_{a}(\Omega)$ is the quantum noise optical transfer function for each signal frequency $\Omega= 2\pi f$, and $\mathcal{T}_{x}(\Omega)$ is the signal optical transfer function. 
 In order to gain a physical picture of the effect, we use a single-mode approximation, which assumes that the signal frequency is well contained within one longitudinal resonance of each cavity.
 The full derivation of the input-output relations and transfer functions is presented in Appendix~\ref{app:subsec:input-output} and transition to the single mode approximation explained in~\ref{app:subsubsec:single-mode}.
 We achieve the QECF regime by setting $\phi=0$ and $\psi=\pi/2$ and also assume the transmissivities of SE and CF mirrors to be very close to each other.
 
 
 

In the single-mode approximation, the noise transfer function and signal transfer function with respect to the GW strain $h(\Omega)=x(\Omega)/L_{\text{arm}}$ are given by
	\begin{equation}
			\mathcal{R}_{a}(\Omega)=\dfrac{\left(\gamma_{cf}-\chi\tilde{\gamma}\right)\Omega+i\left(\Omega^{2}-\omega_{s,\text{\textminus}}^{2}\right)}{\left(\gamma_{cf}+\chi\tilde{\gamma}\right)\Omega-i\left(\Omega^{2}-\omega_{s,\text{+}}^{2}\right)},
			\label{eq:sm-noise-TF}
	\end{equation}
	\begin{equation}
			\mathcal{T}_{h}(\Omega)=
			\dfrac{1}{2}\dfrac{\sqrt{(\chi+1)(1-\chi)}\xi}{\left[\gamma_{cf}+\chi\tilde{\gamma}\right]\Omega-i\left[\Omega^{2}-\omega_{s,\text{+}}^{2}\right]}.
			\label{eq:sm-signal-TF}
	\end{equation}
    We introduced several effective optical parameters that define the behaviour of the system.
    In a standard interferometer, the SEC and arm cavities exchange energy at a specific coupling rate, also called the sloshing frequency:
    	\begin{equation}
		\omega_{s,\text{stand.}}=\sqrt{\dfrac{c^{2}T_{i}^{2}}{4L_{\text{SE}}L_{\textnormal{arm}}}}=\sqrt{\dfrac{\gamma_{\textnormal{arm}}}{\tau_{\textnormal{SE}}}},
		\label{eq:sm-slosh_standard}
	\end{equation}
    where $\gamma_{n} = (T_{n}^2 c)/(4L_n), n\in \{\text{arm},s,cf\}$ are the arm cavity, SEC and CFC bandwidths respectively, with corresponding amplitude transmissivity and cavity length.
    This sloshing frequency is modified by introducing the crystal and the CFC:
	\begin{equation}
		\omega_{s,\text{\textpm}}=\omega_{s,\textnormal{stand.}}\sqrt{\tau_{\text{SE}}\left(\gamma_{s}\dfrac{\tau_{\text{SE}}}{\tau_{cf}}\pm\chi\gamma_{cf}\right)},
		\label{eq:sm-slosh}
	\end{equation}
    where the bandwidth of the interferometer is also modified:
	\begin{equation}
		\tilde{\gamma}=\gamma_{\text{arm}}+\gamma_{s}\dfrac{\tau_{\text{SE}}}{\tau_{cf}}.
		\label{eq:sm-gammatilde}
	\end{equation}
    Both of them depend on the effective gain parameter,
	\begin{equation}
		\chi=\dfrac{e^{2q}-1}{1+e^{2q}},
		\label{eq:gain:chi}
	\end{equation}
 where $q$ is an amplification factor on the single pass through the nonlinear crystal.
The acquired signal strength is proportional to the effective optical power in the arms.
    \begin{equation}
        		\xi=\dfrac{\sqrt{\tau_{\text{SE}}}}{\sqrt{\tau_{\text{arm}}\tau_{cf}}}4Ek_{p}\sqrt{\gamma_{\text{arm}}}\sqrt{\gamma_{s}}\gamma_{cf}L_{\text{arm}},
		\label{eq:sm-xi}
    \end{equation}
	where $E$ is the large classical amplitude of the carrier light field inside the arm cavity and $k_{p}$ is the wave vector of the carrier light field.

    The double-sided spectral density of the output noise is given by the spectral density of the input field $S_\text{in}(\Omega)$ modified by the optical transfer function:
	\begin{equation}
		S_{\rm out}(\Omega)=S_{\rm in}(\Omega)\lvert \mathcal{R}_{a}(\Omega) \rvert^{2},
		\label{noise:spectral:den}
	\end{equation}
 For the purposes of understanding the fundamental behavior of the setup, we assume the incoming light field to be in the vacuum state, $S_\text{in}(\Omega)=1$~\cite{Kimble2000,Braginsky92}, which will be changed to a squeezed state~\cite{Walls1983} in the next section. The strain sensitivity is calculated by comparing the noise power to the response of the readout field to the strain signal,
	\begin{equation}
		S_{h}(\Omega)=\sqrt{\dfrac{S_\text{out}(\Omega)}{\lvert \mathcal{T}_{h}(\Omega) \rvert^{2}}}=\sqrt{\dfrac{\lvert \mathcal{R}_{a}(\Omega) \rvert^{2}}{\lvert \mathcal{T}_{h}(\Omega) \rvert^{2}}}.
		\label{eq:sensitivity}
	\end{equation}
	 From Eqs.\,\ref{eq:sm-noise-TF},\,\ref{eq:sm-signal-TF}, we can compute the strain sensitivity
	\begin{equation}
		S_{h}(\Omega)=2\sqrt{\dfrac{\left(\gamma_{cf}-\chi\tilde{\gamma}\right)^{2}\Omega^{2}+\left(\Omega^{2}-\omega_{s,-}^{2}\right)^{2}}{(\chi+1)(1-\chi)\xi^{2}}}.
		\label{eq:sm-sens}
	\end{equation}

    The behaviour of the QECF can be understood in terms of coupling between the cavities leading to different resonance conditions for the cavities at different frequencies. 
    At low frequencies, the SEC is off resonance (i.e. the field inside is suppressed), and the CFC is on resonance, creating an effective compound mirror with very low reflectivity since the CFC is almost impedance matched, i.e. $T_s\approx T_{cf}$.
    The vacuum field entering the signal port of the interferometer is squeezed very mildly inside the SEC by the nonlinear crystal, since the field is suppressed in the anti-resonant condition.
    The resulting LF sensitivity of QECF resembles the sensitivity of a standard interferometer, with a small suppression of shot noise:
    \begin{equation}
		S_{h}(\Omega)\approx2\Bigl(\gamma_{cf}-\chi\tilde{\gamma}\Bigr)\sqrt{\dfrac{\Omega^{2}+\gamma_{\text{det}}^{2}}{(\chi+1)(1-\chi)\xi^{2}}}
		\label{eq:sm-sens-LF}
	\end{equation}
    with the new detection bandwidth defined as $\gamma_{\text{det}}=\omega_{s,-}^{2}/(\gamma_{cf}-\chi\tilde{\gamma})$.
    
    Without internal squeezing, the LF sensitivity is limited by the detector's bandwidth $\gamma_{\text{baseline}}$:
    \begin{equation}
        \begin{split}
            S_{h}^{\text{baseline}}(\Omega)&=\dfrac{2}{\xi}\sqrt{\gamma_{cf}^{2}\Omega^{2}+\left(\Omega^{2}-\omega_{s,\text{baseline}}^{2}\right)^{2}}\\
            &\approx \dfrac{2 \gamma_{cf}}{\xi}\sqrt{\Omega^{2}+\gamma_{\text{baseline}}^{2}},
        \end{split}
		\label{eq:sm-sens-LF-baseline}
	\end{equation}
    \begin{equation}
		\omega_{s,\text{baseline}}=\omega_{s,\textnormal{stand.}}\tau_{\text{SE}}\sqrt{\dfrac{\gamma_{s}}{\tau_{cf}}},
		\label{eq:sm-slosh-baseline}
	\end{equation}
    \begin{equation}
		\gamma_{\text{baseline}}=\dfrac{\omega_{s,\text{baseline}}^{2}}{\gamma_{cf}}.
		\label{eq:sm-det-baseline}
	\end{equation}
    
    At HF around $\Omega\approx\omega_{s,-}$, the SEC becomes resonant, and the CFC goes off resonance, thus creating a highly reflective compound mirror (similar to the Khalili etalon effect~\cite{Khalili2005}). 
    In this regime, the field inside the SEC is highly amplified, and the nonlinear crystal creates strong squeezing, thus enhancing the sensitivity:
    \begin{equation}
		S_{h}(\Omega)\approx2\dfrac{\left(\gamma_{cf}-\chi\tilde{\gamma}\right)\omega_{s,-}}{\xi\sqrt{(\chi+1)(1-\chi)}}.
		\label{eq:sm-sens-HF}
	\end{equation}
   At the parametric oscillation threshold, $\chi\rightarrow\gamma_{cf}/\tilde{\gamma}$, the sensitivity becomes maximally improved, $S_{h}(\omega_{s,-})\rightarrow 0$.

   If another ratio between the CF and SE mirror transmissivities are chosen (i.e.~they are not close to each other), the QECF loses the ability to generate squeezing simultaneously at LF and HF.

    We note that the resonance condition for the CFC and the SEC depends both on the tuning phases $\psi,\phi$ and on the relation between the transmissivities of the mirrors, $T_{cf}, T_s$ (i.e. the reflection phase for the field $c_s$ off $T_s$). Our choice of $\psi,\phi$ is defined by the requirement of specific conditions to create the desired effect. Other combination of phases are possible: when both the SEC and the CFC are anti-resonant at LF, the effective reflectivity of the CFC is high, and the optical linewidth of the detector is very low; at HF significant squeezing is created, similarly to the QECF. When the SEC is resonant at LF, squeezing is generated at low frequencies too, which is not the case that is relevant to us. Finally, the cavities could also be detuned, thus shifting the optical resonance. This would also create the optical spring effect from quantum radiation pressure, which enhances LF sensitivity~\cite{Braginsky1997,Buonanno2002OPSpring}. This effect can be further enhanced by internal squeezing~\cite{Somiya2016,Korobko2018}. The study of a detuned QECF falls beyond the scope of this paper, where we focus on expanding the bandwidth towards high frequencies.

   \section{\label{sec:Results}Results}
   \subsection{Simultaneous squeezing at low and high frequency}
\begin{figure}
		\includegraphics[width=3.375in]{"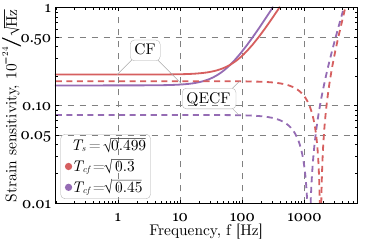"}
		\caption{\label{fig:TEQE-vs-Tt} Strain sensitivity of the Quantum Expander with coherent feedback (QECF), plotted for different transmissivities of the coherent feedback mirror, compared to the coherent feedback detector (CF). The CF shows the sensitivity of the QECF without internally squeezed light. Here, we can see the feature of the QECF to generate simultaneous squeezing at low and high frequency.}
	\end{figure}

    \
    One of the main features of the QECF is its ability to generate simultaneous squeezing at LF and HF despite the presence of detuned cavities. Here, we consider the case without optical losses and QRPN to highlight this feature. \figref{TEQE-vs-Tt} shows the sensitivity of the QECF for a fixed SEM transmissivity of $T_{s}=\sqrt{0.499}$ and different CFM transmissivities. As discussed before, the closer the transmissivities are to each other the closer is the CFC to being impedance matched. At LF, more vacuum field is transmitted through the CFC and squeezed inside the SEC. Thus, the photon shot noise at LF is more squeezed.
    
    \begin{table}[t]
		\caption{\label{tab:parameters:TEQE:QE}Parameters of the Quantum Expander with coherent feedback (QECF) to demonstrate the simultaneous squeezing at low and high frequency.}
		\begin{ruledtabular}
			\begin{tabular}{lc}
				Parameter & QECF
				\\
				\hline
				Laser wavelength 														& $1550\,\text{nm}$\\
				Arm circulating power 													& $4\,\text{MW}$\\
				Test mass weight 														& $200\,\text{kg}$\\
				Input test mass power transmission 										& $7\%$\\
				Signal extraction mirror power transmission	 							& $49.9\%$\\
				Coherent feedback mirror power transmission	 							& $45\%$ or $30\%$\\
				Arm cavity length 														& $20\,\text{km}$\\
				Signal extraction cavity length 										& $56\,\text{m}$\\
				Coherent feedback cavity length 											& $56\,\text{m}$\\
			\end{tabular}
		\end{ruledtabular}
	\end{table}

	\subsection{\label{Res:subsec:TEQEvsQE}QECF compared to the Quantum Expander}

\begin{figure}
		\centering
		\begin{subfigure}{0.99\linewidth}
			\includegraphics[width=\textwidth]{"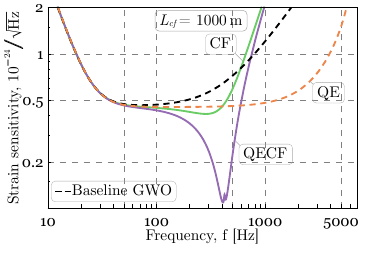"}
            \label{fig:TEQE-vs-QE-Lt1000}
            \vspace{-0.65cm}
		\end{subfigure}
		\begin{subfigure}{0.99\linewidth}
			\includegraphics[width=\textwidth]{"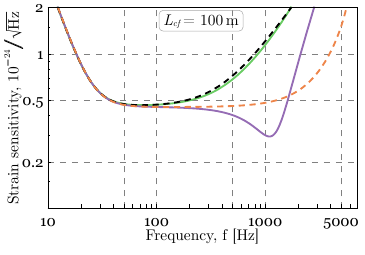"}
            \label{fig:TEQE-vs-QE-Lt100}
            \vspace{-0.65cm}
		\end{subfigure}
		\begin{subfigure}{0.99\linewidth}
			\includegraphics[width=\textwidth]{"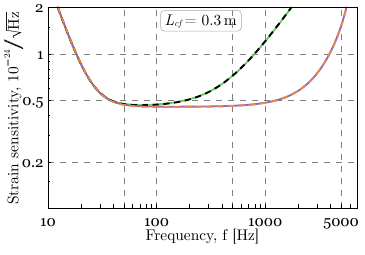"}
            \label{fig:TEQE-vs-QE-Lt0d3}
		\end{subfigure}
		\caption{Demonstration of the flexibility in choice of the high-frequency resonance of the Quantum Expander with coherent feedback (QECF) compared to the Quantum Expander (QE). The sensitivity of the QECF is plotted for different lengths of the coherent feedback cavity. The coherent feedback detector (CF) shows the sensitivity of the QECF without internally squeezed light, while the Baseline GWO is the sensitivity of the QE without internally squeezed light. Both the resonance's frequency and the sensitivity improvement at that frequency scales inversely with the length, disappearing almost entirely in the last plot.}
		\label{fig:TEQE-vs-QE}
	\end{figure}

    Conceptually, the QECF corresponds to the QE with an SE mirror with very high transmissivity, on the order of 95\%, which might pose additional challenges for cavity control.
    The main advantage of the QECF over the QE is the flexibility in choice of the high-frequency resonance where the sensitivity is maximised. \figref{TEQE-vs-QE} shows the sensitivity of QECF plotted for different lengths of the CFC compared to the QE. We can see that increasing its length decreases the detection bandwidth but improves the sensitivity in a certain frequency range, and vice versa. By shortening the CFC, we can obtain the same sensitivity as the QE.
    While the high-frequency resonance of the QE could also be tuned by macroscopically changing the SEC length, in the QECF this effect can be achieved by smaller changes in length, as we show in Appendix \ref{app:subsubsec:scaling-HF}.
    Moreover, tuning the SEC length is potentially challenging in practice due to the crystal and central beam splitter being present inside.
    
	\begin{table}[t]
		\caption{\label{tab:parameters:TEQE:QE}Parameters of the Quantum Expander (QE) and Quantum Expander with coherent feedback (QECF).}
		\begin{ruledtabular}
			\begin{tabular}{lcc}
				Parameter & QE & QECF
				\\
				\hline
				Laser wavelength 														& $1550\,\text{nm}$		& $1550\,\text{nm}$\\
				Arm circulating power 													& $4\,\text{MW}$ 		& $4\,\text{MW}$\\
				Test mass weight 														& $200\,\text{kg}$ 		& $200\,\text{kg}$\\
				Input test mass power transm. 											& $7\%$ 				& $7\%$\\
				End test mass power transm. 											& $5\,\text{ppm}$ 		& $5\,\text{ppm}$\\\noalign{\vskip 1mm} 
                \makecell[l]{Combined power transm. of\\
                the coherent feedback cavity compared to\\
                QE's signal extraction mirror}		                                    & $35\%$			 	& $34.8\%$\\\noalign{\vskip 1mm} 
				Signal extraction mirror power transm.	 								& $35\%$ 				& $29.9\%$\\
				Coherent feedback mirror power transm.		 							& -			 			& $3.7\%$\\
				Arm cavity length 														& $20\,\text{km}$ 		& $20\,\text{km}$\\
				Signal extraction cavity length 										& $56\,\text{m}$ 		& $56\,\text{m}$\\
				Coherent feedback cavity length 											& -				 		& $100\,\text{m}$\\
				Loss inside the signal extraction cavity					 			& $1500\,\text{ppm}$ 	& $1500\,\text{ppm}$\\
				Photodiode detection efficiency 										& $99.5\%$ 				& $99.5\%$\\
			\end{tabular}
		\end{ruledtabular}
	\end{table}
	
	\subsection{\label{Res:subsec:TEQEvsNEMO}QECF compared to NEMO}
\begin{figure}
		\includegraphics[width=3.375in]{"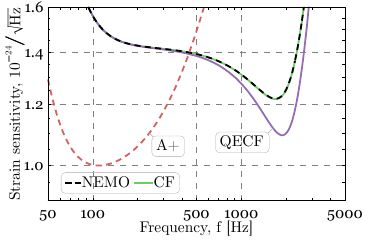"}
		\caption{\label{fig:TEQE-NEMO-short} Strain sensitivity of the Quantum Expander with coherent feedback with internal squeezing (QECF) and without it (CF), compared to NEMO. Parameters of CF are chosen to follow the sensitivity of NEMO to demonstrate sensitivity enhancement in the presence of internal squeezing in QECF. A+~\cite{Barsotti2018} sensitivity is plotted for the reference.}
	\end{figure}

    The QECF features the high-frequency resonance with a target of observing neutron-star mergers. 
    This relates it to the proposed NEMO detector, which targets the same sensitivity range.
    We use NEMO design as a benchmark for our sensitivity optimization.
    First, in \figref{TEQE-NEMO-short} we compare the two designs, matching the parameters of the coherent feedback detector (CF), which corresponds to the QECF without internal squeezing, to NEMO as close as possible to follow the sensitivity of NEMO. We show that we can obtain the same sensitivity as NEMO with the CF.
    While NEMO plans for $354\,\text{m}$ long SEC, the combined length of the SEC and the CFC in the QECF and CF is 82.5\,m ($56\,\text{m}$ and $26.5\,\text{m}$ respectively).
    This potentially reduces costs and simplifies experimental setup and operation of the detector. Furthermore, the multiple cavity structure makes it easier to tune the CF sensitvity. The LF and peak sensitivity can be improved but at the expanse of a smaller HF resonance.

\begin{figure}
		\includegraphics[width=3.375in]{"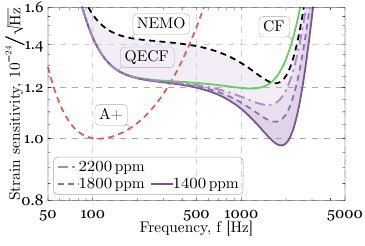"}
		\caption{\label{fig:TEQE-NEMO-loss} Strain sensitivity of the Quantum Expander with coherent feedback (QECF), plotted for different amounts of loss inside the signal extraction cavity $\lambda_{s}$, compared to NEMO. The coherent feedback detector (CF) shows the sensitivity of the QECF without internally squeezed light and for the same loss inside the signal extraction cavity as NEMO. A+ sensitivity is plotted for the reference.}
	\end{figure}

    The CFC gives freedom in optimizing the cavity lengths and reflectivities of the mirrors while maintaining the same HF resonance frequency.
    Such optimization allows to enhance the sensitivity compared to NEMO in a broad band, as we show in \figref{TEQE-NEMO-loss}.
    For this optimization we also take into account that internal loss inside the SEC is increased compared to NEMO due to addition of the crystal and associated bulk material absorption and reflections at the sides of the crystal.
    To account for these additional losses, we increase the intra-cavity loss from 1400\,ppm assumed by NEMO by 400 and 800\,ppm.
    A loss estimate comes from the known value for the material absorption on the level of 10\,ppm/cm~\cite{Ast2011} and anti-reflective coating with reflectivity of 100\,ppm~\cite{Mueller2016}. We assume that the crystal is only a few millimeters in size and thus, that we can neglect the loss due to absorption, which would result in $\approx400\,$ppm loss in double pass.
    We leave room for additional losses and show that in the worst-case scenario the QECF still has the capability to outperform NEMO in a broad band. Of course, the internal squeezing approach can be used in NEMO as well~\cite{adya2020quantum}, but we don't discuss this case in detail here since we could achieve the resulting sensitivity with the QECF as well.
 
\begin{figure}
 		\begin{subfigure}{0.9\linewidth}
			\includegraphics[width=\textwidth]{"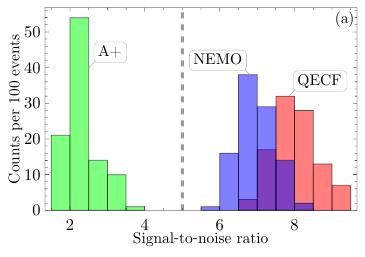"}
            \phantomsubcaption
			\label{fig:AstroAnalysis-TEQE-NEMO-Aplus}
                \vspace{-0.65cm}
		\end{subfigure}
		\begin{subfigure}{0.9\linewidth}
			\includegraphics[width=\textwidth]{"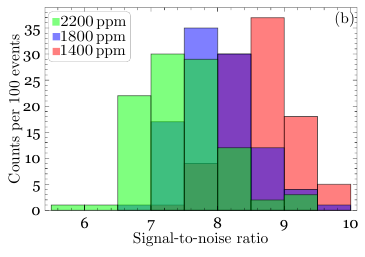"}
            \phantomsubcaption
			\label{fig:AstroAnalysis-TEQE-different-losses}
		\end{subfigure}
		\caption{\label{fig:AstroAnalysis} Histogram for the SNR of the loudest events for 100 Monte-Carlo simulations of neutron star mergers in the kHz GW regime. (a) shows the results of one realization where we compared the Quantum Expander with coherent feedback (QECF), NEMO and A+. (b) shows the results of another realization where we compared QECF for different amounts of loss inside the signal extraction cavity $\lambda_{s}$.}
	\end{figure}

     In \figref{AstroAnalysis}, we demonstrate the improved detection rates of neutron star mergers. We simulated one year of observation of kHz GWs ($1$\textminus$4\,\text{kHz}$) from neutron star mergers~\cite{Yang2018Evo,Bauswein2012} by running Monte-Carlo simulations for different source parameters and a given equation of state (see the details in the Appendix~\ref{app:subsec:astro-analyis}).
     As we can see in \figref{AstroAnalysis-TEQE-NEMO-Aplus}, both NEMO and the QECF significantly improve the detection probability of kHz GWs compared to A+. Here, we assume the detection threshold for a GW event to be a signal-to-noise ratio (SNR) of $5$. While no event has an SNR above the threshold for A+, all events have an SNR above the threshold for the QECF and NEMO.
     The increased sensitivity of the QECF improves the median of the SNRs from $6.94$ (NEMO) to $7.98$ (QECF). This improvement also depends on the actual amount of loss inside the SEC, which we demonstrated in \figref{AstroAnalysis-TEQE-different-losses}.

    Since the QECF improves the sensitivity at low frequencies as well, it also enhances the cosmological reach for binary black hole mergers by a factor $\approx 1.5$ in the volume of the Universe. It is worth mentioning that improved SNRs alone won't make a big difference. A good sky localisation of the source is still needed to boost the multimessenger astronomy~\cite{Gardner2023,Gupta2024}.

	\begin{table*}[t]
		\caption{\label{tab:parameters:TEQE:NEMO} Parameters NEMO, coherent feedback detector (CF) and Quantum Expander with coherent feedback (QECF).}
		\begin{ruledtabular}
			\begin{tabular}{lcccc}
				Parameter & NEMO & CF & \makecell{QECF\\matched to NEMO} & \makecell{QECF\\optimized}
				\\
				\hline
				Laser wavelength 											& $2000\,\text{nm}$		& $2000\,\text{nm}$	& $2000\,\text{nm}$ 	& $2000\,\text{nm}$\\
				Arm circulating power 										& $4.5\,\text{MW}$ 		& $4.5\,\text{MW}$ 	& $4.5\,\text{MW}$ 		& $4.5\,\text{MW}$\\
				Test mass weight 											& $74.1\,\text{kg}$ 	& $74.1\,\text{kg}$ & $74.1\,\text{kg}$ 	& $74.1\,\text{kg}$\\
				Input test mass power transmission 							& $1.4\%$ 				& $1.4\%$ 			& $1.4\%$ 				& $1.4\%$\\
				End test mass power transmission 							& $5\,\text{ppm}$ 		& $5\,\text{ppm}$ 	& $5\,\text{ppm}$ 		& $5\,\text{ppm}$\\\noalign{\vskip 1mm}
                \makecell[l]{Combined power transmission of the coherent feedback cavity\\
                compared to NEMO's signal extraction mirror}	            & $4.8\%$			 	& $4.8\%$		 	& $4.8\%$ 				& $6.36\%$\\\noalign{\vskip 1mm}
				Signal extraction mirror power transmission 				& $4.8\%$ 				& $29.9\%$		 	& $29.9\%$ 				& $46\%$ \\
				Coherent feedback mirror power transmission 					& -			 			& $0.435\%$		 	& $0.435\%$ 			& $1\%$\\
				Arm cavity length 											& $4\,\text{km}$ 		& $4\,\text{km}$ 	& $4\,\text{km}$ 		& $4\,\text{km}$\\
				Signal extraction cavity length 							& $354\,\text{m}$ 		& $56\,\text{m}$ 	& $56\,\text{m}$ 		& $56\,\text{m}$\\
				Coherent feedback cavity length 								& -				 		& $26.5\,\text{m}$ 	& $26.5\,\text{m}$		& $45\,\text{m}$\\
				Loss inside the signal extraction cavity without crystal 	& $1400\,\text{ppm}$ 	& $1400\,\text{ppm}$& $1400\,\text{ppm}$ 	& $1400\,\text{ppm}$\\
				Additional loss due to the crystal 							& -	 					& -		 			& $400\,\text{ppm}$ 	& $400\,\text{ppm}$\\
				Photodiode detection efficiency 							& $99.7\%$ 				& $99.7\%$ 			& $99.7\%$ 				& $99.7\%$\\
				Detected external squeezing 						& $7\,\text{dB}$ 		& $7\,\text{dB}$ 	& $7\,\text{dB}$ 		& $7\,\text{dB}$\\
			\end{tabular}
		\end{ruledtabular}
	\end{table*}

	\section{\label{sec:Conclusion}Summary and conclusion}

    Coherent feedback is commonly used in the context of quantum networks for modifying the response of the plants generating quantum correlated states.
    In our work we used this approach for gravitational wave detection, modifying the spectral shape of the quantum noise in the detector.
    We demonstrated the benefits of QECF for detecting high-frequency GW signal in a realistic setting, where the detection and intra-cavity losses limit the impact of quantum enhancement~\cite{Miao2019LOD}.
    Our analysis of different cases, including the added loss due to the nonlinear crystal inside the detector,
    demonstrated the broadband enhancement to the sensitivity of a HF detector, compared to NEMO or QE. The demonstrated gain in sensitivity, which corresponds to an increase in light power inside NEMO's arm cavities by about $32\%$, significantly expands the volume in which binary black hole mergers are detectable by a factor $\approx 1.5$. We believe that increased complexity due to the additional mirror and nonlinear crystal is within current technological capabilities. The intra-cavity loss, which is one of the main limitations of quantum enhancement, is not greatly increased by the additional nonlinear crystal according to our estimates. However, even in the case of a significant added intra-cavity loss QECF still provides the sensitivity gain. The detailed implementation of the crystal inside the detector shall remain the goal of future studies.
    Other technical challenges, associated with phase noise~\cite{Schnabel2017,Kijbunchoo2020}, dephasing~\cite{Kwee2006}, mode mismatch~\cite{McCuller2021, goodwin2023transverse} and parametric amplification process will need to be addressed.

     Our new design demonstrates the promise of using coherent feedback cavities with quantum enhancement for flexible tuning of the detector sensitivity. Such tuning allows to optimize the detector between the observation runs to enhance its sensitivity to particular GW sources, based on the advances in understanding of their properties.
     Our approach can extend beyond gravitational-wave detection towards other large-scale dark matter or gravity sensors~\cite{Backes2021,Carney2021}.
	\begin{acknowledgments}
    We thank Y. Ma for fruitful discussions. This work was supported by the Deutsche Forschungsgemeinschaft (DFG) under Germany’s Excellence Strategy EXC 2121 “Quantum Universe”-390833306.
	\end{acknowledgments}
		
	\appendix
	\section{Appendices}
	
	\subsection{\label{app:subsec:input-output}Input-Output relations}
    \begin{figure*}
		  \includegraphics[width=5.0625in]{"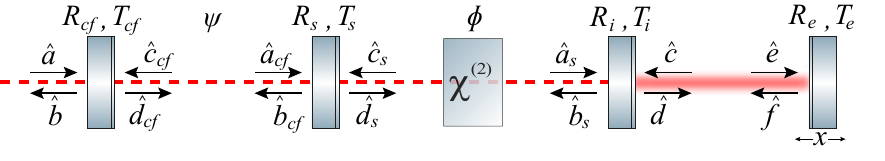"}
		  \caption{\label{fig:TEQE-setup-theory_w_fields}Simplified theoretical setup for the Quantum Expander with coherent feedback. $R_{cf,s,i,e}$ and $T_{cf,s,i,e}$ are amplitude reflectivities and transmissivities of the coherent feedback, signal extraction, input and end mirrors respectively. $\psi$ is the single trip tuning of the coherent feedback cavity and $\phi$ is the single trip tuning of the signal extraction cavity. x is a small mirror displacement due to a gravitational wave.}
	\end{figure*}
	The steady-state input-output relations of the phase quadrature of the light field of the QECF are as follows:
	\allowdisplaybreaks
	\begin{subequations}
		\label{eq:input-output}
		\begin{align}
			&\hat{c}=\hat{d} e^{2i\Omega\tau_{\text{arm}}}+2ik_{p}E x e^{i\Omega\tau_{\text{arm}}},
			\label{eq:c}
			\\
			&\hat{d}=T_{i}\hat{a}_{s}+R_{i}\hat{c},
			\label{eq:d}
			\\
			&\hat{a}_{s}=\hat{d}_{s}e^{-q}e^{i\phi}e^{i\Omega\tau_{\text{SE}}},
			\label{eq:as}
			\\
			&\hat{b}_{s}=-R_{i}\hat{a}_{s}+T_{i}\hat{c},
			\label{eq:bs}
			\\
			&\hat{c}_{s}=\hat{b}_{s}e^{-q}e^{i\phi}e^{i\Omega\tau_{\text{SE}}},
			\label{eq:cs}
			\\
			&\hat{d}_{s}=T_{s}\hat{a}_{cf}+R_{s}\hat{c}_{s},
			\label{eq:ds}
			\\
			&\hat{a}_{cf}=\hat{d}_{cf}e^{i\psi}e^{i\Omega\tau_{cf}},
			\label{eq:a}
			\\
			&\hat{b}_{cf}=-R_{s}\hat{a}_{cf}+T_{s}\hat{c}_{s},
			\label{eq:btr}
			\\
			&\hat{c}_{cf}=\hat{b}_{cf}e^{i\psi}e^{i\Omega\tau_{cf}},
			\label{eq:ctr}
			\\
			&\hat{d}_{cf}=T_{cf}\hat{a}+R_{cf}\hat{c}_{cf},
			\label{eq:dtr}
			\\
			&\hat{b}=-R_{cf}\hat{a}+T_{cf}\hat{c}_{cf},
			\label{eq:b}
		\end{align}
	\end{subequations}
	where $R_{i,s,cf}$, $T_{i,s,cf}$ are the amplitude reflectivity and transmissivity of the input, signal extraction and coherent feedback mirror; $q$ is an amplification factor on the single pass through the crystal; $\tau_{\text{arm},\text{SE},cf}=L_{\text{arm},\text{SE},cf}/c$ is the single trip time in the arm cavity of length $L_{\text{arm}}$, SEC of length $L_{\text{SE}}$ and CFC of length $L_{cf}$, with $c$ being the speed of light; the arm cavity is resonant at zero signal frequency ($\Omega=0$), $\phi$ is the single trip tuning of the SEC and $\psi$ is the single trip tuning of the CFC; $x$ is a small displacement of the end mirror due to the GW signal; $E$ is the large classical amplitude of the carrier light field inside the arm cavity and $k_{p}$ is the wave vector of the carrier light field.
 
	By solving Eqs. \ref{eq:input-output}, we get an expression for the output $\hat{b}$, which can be split into a noise part and signal part, where $\mathcal{R}_{a}(\Omega)$ is the noise optical transfer function and $\mathcal{T}_{x}(\Omega)$ is the signal optical transfer function:
	\begin{equation}
		\hat{b}=\mathcal{R}_{a}(\Omega)\hat{a}+\mathcal{T}_{x}(\Omega)x,
		\label{app:eq:bsol}
	\end{equation}
	\begin{multline}
		\mathcal{R}_{a}(\Omega)=\dfrac{e^{2q}\left(-1+e^{2i\Omega\tau_{\text{arm}}}R_{i}\right)\left(R_{cf}+e^{2i(\psi+\Omega\tau_{cf})}R_{s}\right)}{c_{\text{den}}}\\+\dfrac{e^{2i(\phi+\Omega\tau_{\text{SE}})}\left(e^{2i\Omega\tau_{\text{arm}}}-R_{i}\right)\left(e^{2i(\psi+\Omega\tau_{cf})}+R_{cf}R_{s}\right)}{c_{\text{den}}},
		\label{eq:noise-TF}
	\end{multline}
	\begin{equation}
		\mathcal{T}_{x}(\Omega)= -\dfrac{2ie^{q+i(\phi+\psi+\Omega(\tau_{\text{arm}}+\tau_{\text{SE}}+\tau_{cf}))}E k_{p}T_{i}T_{s}T_{cf}}{c_{\text{den}}},
		\label{eq:signal-TF}
	\end{equation}
	where we defined
	\begin{multline}
		c_{\text{den}}=e^{2i(\phi+\Omega\tau_{\text{SE}})}\Bigl(e^{2i\Omega\tau_{\text{arm}}}-R_{i}\Bigr)\Bigl(e^{2i(\psi+\Omega\tau_{cf})}R_{cf}+R_{s}\Bigr)\\
		+e^{2q}\Bigl(-1+e^{2i\Omega\tau_{\text{arm}}}R_{i}\Bigr)\Bigl(1+e^{2i(\psi+\Omega\tau_{cf})}R_{cf}R_{s}\Bigr).
		\label{eq:denom}
	\end{multline}
     Both transfer functions depend, amongst other parameters, on the properties of the cavities. 

    \subsubsection{\label{app:subsubsec:single-mode} Single mode approximation}
    
    The expressions of the noise and signal transfer functions~(Eqs.\,\ref{eq:noise-TF},\,\ref{eq:signal-TF}) can be simplified by doing a single-mode-approximation, which assumes that the signal frequency of interest is well contained within one free spectral range of the longitudinal resonances of each cavity. Therefore, we assume $\Omega\tau_{\text{arm},\text{SE},cf}\ll1$ and $T_{i,s,cf}\ll1$, so that $e^{i\Omega\tau_{\text{arm},\text{SE},cf}}\approx 1+i\Omega\tau_{\text{arm},\text{SE},cf}$ and $R_{i,s,cf}\approx 1-T_{i,s,cf}^{2}/2$. The assumptions made simplify the numerator and denominator of the transfer functions to polynomials of $\Omega$. The coefficients of the polynomials can be further simplified by dropping higher order terms until only the lowest order term or two lowest order terms are left. We monitored the correctness of this method by comparing the roots of the numerator of the noise transfer function with the high-frequency resonance of the actual solution since they should be identical. In this regime, the noise transfer function and signal transfer function with respect to the GW strain $h(\Omega)=x(\Omega)/L_{\text{arm}}$ are given by~Eqs.\,\ref{eq:sm-noise-TF},\,\ref{eq:sm-signal-TF}.
    
    \subsubsection{\label{app:subsubsec:scaling-HF} Scaling of the high-frequency resonance}
    
\begin{figure}
		\includegraphics[width=3.375in]{"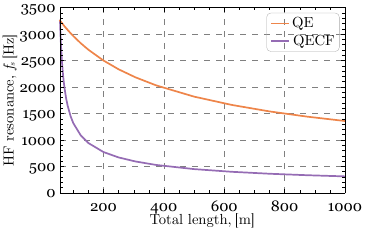"}
		\caption{\label{fig:HF-resonance-scaling} Scaling of the high-frequency resonance versus total length of the extraction cavities. For the Quantum Expander (QE) we changed the length of the signal extraction cavity, while for the Quantum Expander with coherent feedback (QECF) we kept the length of the signal extraction cavity constant ($56\,\text{m}$) and only changed the length of the coherent feedback cavity. The plot highlights the more efficient tuning of the high-frequency resonance in the QECF compared to the QE.}
    \end{figure}

    \figref{HF-resonance-scaling} compares the efficiency of tuning the high-frequency resonance of the QECF and QE. Here, we determined the high-frequency resonance of both systems without optical losses and QRPN. For the QE we changed the length of the signal extraction cavity, while for the QECF we kept the length of the signal extraction cavity constant ($56\,\text{m}$) and only changed the length of the coherent feedback cavity. The efficiency of the tuning of the QECF is significantly better, especially between $56\,\text{m}$ and $\sim400\,\text{m}$.

    \subsection{\label{app:subsec:astro-analyis}Astrophysical analysis}
        Here we present the details of the astrophysical analysis. The analysis focuses on binary neutron star mergers and follows the method described in \cite{Yang2018,Miao2018}. The event rate is estimated to be $1\,\text{Mpc}^{-3}\,\text{Myr}^{-1}$ and the searching distance to be $1\,\text{Gpc}$. We ran 100 Monte-Carlo simulations each with 1000 samples. Based on the event rate and search distance, this corresponds to one year of observation. The mass of each neutron star is assumed to follow a Gaussian distribution centered at $1.33\,\text{M}_{\odot}$ with variance $0.09\,\text{M}_{\odot}$. Here, we exclude binary neutron stars with a total mass larger than $3.45\,\text{M}_{\odot}$ because they will quickly collapse into a black hole. The sky position, inclination and polarization angles are taken into account by multiplying the gravitational wave amplitude (Eq. \ref{eq:analysis-amplitude}) by a sky-averaged reduction factor~\cite{Yang2018}. The distribution of the source distance follows a uniform distribution between $50\,\text{Mpc}$ and $1000\,\text{Mpc}$~\cite{Miao2018}. We approximate the post-merger waveform as the damped oscillation of a single mode, the parameters of which depend on the neutron-star equation of state. The amplitude in the frequency domain is given by:
	\begin{equation}
		h(f)=\dfrac{2}{5}\dfrac{50}{\pi\,\text{d}}h_{p}\dfrac{Q\Bigl[2 f_{p}Q\cos(\phi_{0})-[f_{p}-2ifQ]\sin(\phi_{0})\Bigr]}{f_{p}^{2}-4iff_{p}Q-4Q^{2}(f^{2}-f_{p}^{2})},
		\label{eq:analysis-amplitude}
	\end{equation}
	where $d$ is the distance to the source in Mpc, $h_{p}$ is the peak value of the wave amplitude, $Q$ is the quality factor of the post-merger oscillation, $\phi_{0}$ is the initial phase of the source, $f_{p}$ is the peak frequency of the waveform and $2/5$ is the sky-averaged reduction factor for a L-shaped GW detector. The equation of state of the neutron star is assumed to be relatively stiff \cite{Shen1998} and therefore, $Q=23.3$ and $h_{p}=5\times10^{-22}$. The peak frequency is given by:
	\begin{equation}
		f_{p}=1\,\text{kHz}\left(\dfrac{m_{1}+m_{2}}{\text{M}_{\odot}}\right)\left[a_{2}\left(\dfrac{R}{1\,\text{km}}\right)^{2}+a_{1}\dfrac{R}{1\,\text{km}}+a_{0}\right],
		\label{eq:analysis-peak}
	\end{equation}
	where $R=14.42\,\text{km}$ is the radius of each neutron star with masses $m_{1,2}$ and the parameters $a_{2}=0.0157$, $a_{1}=-0.5495$, $a_{0}=5.503$ \cite{Bauswein_2016}. The signal-to-noise ratio is defined as:
	\begin{equation}
		\text{SNR}=2\sqrt{\int_{f_{\text{min}}}^{f_{\text{max}}}df\,\dfrac{\vert h(f) \vert^{2}}{S_{hh}(f)}},
	\end{equation}
	where $S_{hh}(f)$ is the single-sided noise spectral density of the detectors and $f_{\text{min}}=1\,\text{kHz}$ and $f_{\text{max}}=4\,\text{kHz}$ are the integration limits. For each Monte-Carlo simulation, we selected out the loudest event.

	\newpage
	\onecolumngrid
	
	\section*{Supplementary: Full calculation of the strain sensitivity}
    
	Optical loss is a huge problem for gravitational wave observatories (GWOs). Any source of optical loss reduces the sensitivity of a GWO due to the mixing with vacuum. In this section, we compute the strain sensitivity including optical losses, radiation pressure and injection of external squeezed light by using a transfer matrix approach. The  following calculations mirror those in \cite{Korobko2020}.
	\subsubsection{\label{app:subsubsec:loss}Input-output relation}
	\begin{figure}[!b]
		\includegraphics[width=5.0625in]{"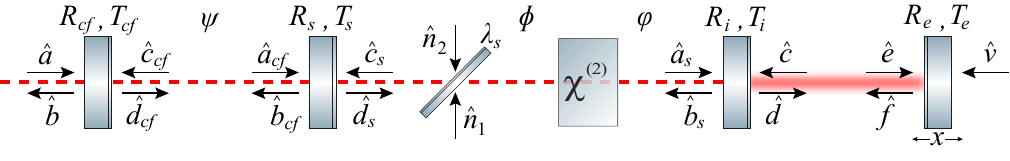"}
		\caption{\label{fig:TEQE_setup_loss} Quantum fields inside the Quantum Expander with coherent feedback. $R_{cf,s,i,e}$ and $T_{cf,s,i,e}$ are amplitude reflectivities and transmissivities of the coherent feedback, signal extraction, input and end test mirror. $\psi$ is the phase delay due to the detuning of the coherent feedback cavity. $\phi$ and $\varphi$ are the phase delays due to the signal extraction cavity detuning before and after the crystal. Intra-cavity loss is represented by a beamsplitter with power reflectivity $\lambda_{s}$.}
	\end{figure}
	The setup of the Quantum Expander with coherent feedback (QECF) and quantum fields inside the system with losses are shown in Fig. \ref{fig:TEQE_setup_loss}.\\ We define a vector $\hat{a}(\Omega)=\left\{\hat{a}_{\text{phase}}(\Omega),\hat{a}_{\text{amplitude}}(\Omega)\right\}^{\text{T}}$ to describe the phase and amplitude quadrature. We will start by writing down the input-output relations and solving them. The coherent feedback and signal extraction cavity can rotate the quadratures due to its detuning from resonance. The parametric amplification process squeezes and rotates the quadratures. We introduce a rotation matrix:
	\begin{equation}
		\mathbb{O}(\phi)=
		\begin{pmatrix}
			\cos(\phi) & -\sin(\phi)\\
			\sin(\phi) & \cos(\phi)
		\end{pmatrix}, \forall \phi
		\label{eq:loss-rot-matrix}
	\end{equation}
	and squeezing matrix:
	\begin{equation}
		\mathbb{S}=
		\begin{pmatrix}
			e^{q} & 0\\
			0 & e^{-q}
		\end{pmatrix},
		\label{eq:loss-sq-matrix}
	\end{equation}
	where $q$ is an amplification factor on the single pass through the crystal.\\
	The equations for the coherent feedback cavity are as follows:
	\begin{subequations}
		\begin{align}
			&\hat{a}_{cf}=\mathbb{N}(\psi)\hat{d}_{cf},\\
			&\hat{b}_{cf}=-R_{s}\hat{a}_{cf}+T_{s}\hat{c}_{s},\\
			&\hat{c}_{cf}=\mathbb{N}(\psi)\hat{b}_{cf},\\
			&\hat{d}_{cf}=T_{cf}\hat{a}+R_{cf}\hat{c}_{cf},
		\end{align}
		\label{eq:loss-tr-relations}
	\end{subequations}
	where we have defined
	\begin{equation}
		\mathbb{N}(\psi)=\mathbb{O}(\psi)e^{i\Omega\tau_{cf}}
		\label{eq:loss-N}
	\end{equation}
	and $R_{cf,s},T_{cf,s}$ are the amplitude reflectivities and transmissivities of the coherent feedback and signal extraction mirror, $\tau_{cf}=L_{cf}/c$ is the coherent feedback cavity global delay and $\psi$ is the phase delay due to the cavity detuning. The system of equations Eqs. \ref{eq:loss-tr-relations} can be solved and we find the following solutions:
	\begin{subequations}
		\begin{align}
			&\hat{a}_{cf}=\mathbb{N}(\psi)\mathbb{A}\left(T_{cf}\hat{a}+R_{cf}T_{s}\mathbb{N}(\psi)\hat{c}_{s}\right),\\
			&\hat{b}_{cf}=\mathbb{A}\left(-R_{s}T_{cf}\mathbb{N}(\psi)\hat{a}+T_{s}\hat{c}_{s}\right),\\
			&\hat{c}_{cf}=N(\psi)\mathbb{A}\left(-R_{s}T_{cf}\mathbb{N}(\psi)\hat{a}+T_{s}\hat{c}_{s}\right),\\
			&\hat{d}_{cf}=\mathbb{A}\left(T_{cf}\hat{a}+R_{cf}T_{s}\mathbb{N}(\psi)\hat{c}_{s}\right),
		\end{align}
		\label{eq:loss-tr-sol}
	\end{subequations}
	where we have defined
	\begin{equation}
		\mathbb{A}=\left(\mathbb{I}+R_{cf}R_{s}\mathbb{N}^{2}(\psi)\right)^{-1}.
		\label{eq:loss-A}
	\end{equation}
	The solutions can be used to obtain the input-output relations for the coherent feedback cavity:
	\begin{subequations}
		\begin{align}
			&\hat{b}=-R_{cf}\hat{a}+T_{cf}\hat{c}_{cf}=-\tilde{\mathbb{R}}_{b}\hat{a}+\tilde{\mathbb{T}}_{b}\hat{c}_{s},\\
			&\hat{d}_{s}=T_{s}\hat{a}_{cf}+R_{s}\hat{c}_{s}=\tilde{\mathbb{R}}_{d}\hat{c}_{s}+\tilde{\mathbb{T}}_{d}\hat{a},\label{eq:loss-tr-input-output-b}
		\end{align}
		\label{eq:loss-tr-input-output}
	\end{subequations}
	where
	\begin{subequations}
		\begin{align}
			&\tilde{\mathbb{R}}_{b}=R_{cf}\mathbb{I}+R_{s}T_{cf}^{2}\mathbb{N}(\psi)\mathbb{A}\mathbb{N}(\psi),\\
			&\tilde{\mathbb{R}}_{d}=R_{s}\mathbb{I}+R_{cf}T_{s}^{2}\mathbb{N}(\psi)\mathbb{A}\mathbb{N}(\psi),\\
			&\tilde{\mathbb{T}}_{b}=T_{cf}T_{s}\mathbb{N}(\psi)\mathbb{A},\\
			&\tilde{\mathbb{T}}_{d}=T_{s}T_{cf}\mathbb{N}(\psi)\mathbb{A}
		\end{align}
		\label{eq:loss-tr-TF}
	\end{subequations}
	are the transfer matrices for the fields.\\
	For the signal extraction cavity, we get the following set of equations:
	\begin{subequations}
		\begin{align}
			&\hat{a}_{s}=\sqrt{1-\lambda_{s}}\mathbb{M}\left[\varphi,\phi\right]\hat{d}_{s}+\sqrt{\lambda_{s}}\mathbb{M}\left[\varphi,\phi\right]\hat{n}_{1},\\
			&\hat{b}_{s}=-R_{i}\hat{a}_{s}+T_{i}\hat{c},\\
			&\hat{c}_{s}=\sqrt{1-\lambda_{s}}\mathbb{M}\left[\phi,\varphi\right]\hat{b}_{s}-\sqrt{\lambda_{s}}\hat{n}_{2},\\
			&\hat{d}_{s}=T_{s}\hat{a}_{cf}+R_{s}\hat{c}_{s},
		\end{align}
		\label{eq:loss-sq-relations}
	\end{subequations}
	where we have defined
	\begin{equation}
		\mathbb{M}\left[\varphi,\phi\right]=\mathbb{O}(\varphi)\mathbb{O}(\theta)\mathbb{S}\mathbb{O}^{\dagger}(\theta)\mathbb{O}(\phi)e^{i\Omega\tau_{\text{SE}}}
		\label{eq:loss-M}
	\end{equation}
	and $R_{i},T_{i}$ are the amplitude reflectivity and transmissivity of the input test mirror, $\tau_{\text{SE}}=L_{\text{SE}}/c$ is the signal extraction cavity global delay, $\phi$ and $\varphi$ are the phase delays due to the cavity detuning before and after the crystal, $\lambda_{s}$ is the power loss inside the signal extraction cavity and $\theta$ is the squeezing angle.\\
	The next step is to insert $\hat{d}_{s}$ from Eq. \ref{eq:loss-tr-input-output-b} into the Eqs. \ref{eq:loss-sq-relations}. We obtain the input-output relations for the system, which consists of the coherent feedback and signal extraction cavity:
	\begin{subequations}
		\begin{align}
			&\hat{b}=-\grave{\mathbb{R}}_{b}\hat{a}+\grave{\mathbb{T}}_{b}\hat{c}+\grave{\mathcal{L}}_{b1}\hat{n}_{1}+\grave{\mathcal{L}}_{b2}\hat{n}_{2},\\
			&\hat{d}=\grave{\mathbb{R}}_{d}\hat{c}+\grave{\mathbb{T}}_{d}\hat{a}+\grave{\mathcal{L}}_{d1}\hat{n}_{1}+\grave{\mathcal{L}}_{d2}\hat{n}_{2},\label{eq:loss-tr-sq-input-output-b}
		\end{align}
		\label{eq:loss-tr-sq-input-output}
	\end{subequations}
	where we have defined the following transfer matrices
	\begin{subequations}\label{loss:addc:sq:TF}
		\begin{align}
			&\grave{\mathbb{R}}_{b}=\tilde{\mathbb{R}}_{b}+R_{i}(1-\lambda_{s})\tilde{\mathbb{T}}_{b}\mathbb{M}\left[\phi,\varphi\right]\mathbb{G}_{a}\mathbb{M}\left[\varphi,\phi\right]\tilde{\mathbb{T}}_{d},\\
			&\grave{\mathbb{R}}_{d}=R_{i}\mathbb{I}+T_{i}^{2}(1-\lambda_{s})\mathbb{G}_{a}\mathbb{M}\left[\varphi,\phi\right]\tilde{\mathbb{R}}_{d}\mathbb{M}\left[\phi,\varphi\right],\\
			&\grave{\mathbb{T}}_{b}=\sqrt{1-\lambda_{s}}\tilde{\mathbb{T}}_{b}\mathbb{M}\left[\phi,\varphi\right]\left(T_{i}\mathbb{I}-R_{i}T_{i}(1-\lambda_{s})\mathbb{G}_{a}\mathbb{M}\left[\varphi,\phi\right]\tilde{\mathbb{R}}_{d}\mathbb{M}\left[\phi,\varphi\right]\right),\\
			&\grave{\mathbb{T}}_{d}=T_{i}\sqrt{1-\lambda_{s}}\mathbb{G}_{a}\mathbb{M}\left[\varphi,\phi\right]\tilde{\mathbb{T}}_{d},\\
			&\grave{\mathcal{L}}_{b1}=-R_{i}\sqrt{1-\lambda_{s}}\sqrt{\lambda_{s}}\tilde{\mathbb{T}}_{b}\mathbb{M}\left[\phi,\varphi\right]\mathbb{G}_{a}\mathbb{M}\left[\varphi,\phi\right],\\
			&\grave{\mathcal{L}}_{d1}=T_{i}\sqrt{\lambda_{s}}\mathbb{G}_{a}\mathbb{M}\left[\varphi,\phi\right],\\
			&\grave{\mathcal{L}}_{b2}=R_{i}(1-\lambda_{s})\sqrt{\lambda_{s}}\tilde{\mathbb{T}}_{b}\mathbb{M}\left[\phi,\varphi\right]\mathbb{G}_{a}\mathbb{M}\left[\varphi,\phi\right]\tilde{\mathbb{R}}_{d}-\sqrt{\lambda_{s}}\tilde{\mathbb{T}}_{b},\\
			&\grave{\mathcal{L}}_{d2}=-T_{i}\sqrt{1-\lambda_{s}}\sqrt{\lambda_{s}}\mathbb{G}_{a}\mathbb{M}\left[\varphi,\phi\right]\tilde{\mathbb{R}}_{d},\\
			&\mathbb{G}_{a}=\left(\mathbb{I}+R_{i}(1-\lambda_{s})\mathbb{M}\left[\varphi,\phi\right]\tilde{\mathbb{R}}_{d}\mathbb{M}\left[\phi,\varphi\right]\right)^{-1}.
		\end{align}
	\end{subequations}
	The equations for the arm cavity are as follows:
	\begin{subequations}
		\begin{align}
			&\hat{e}=\mathbb{P}(\delta)\hat{d},\\
			&\hat{f}=R_{e}\hat{e}+T_{e}\hat{v}+2k_{p}R_{e}\mathbb{O}(\pi/2)E\hat{x}(\Omega),\\
			&\hat{c}=\mathbb{P}(\delta)\hat{f},\\
			&\hat{d}=R_{i}\hat{c}+T_{i}\hat{a}_{s},
		\end{align}
		\label{eq:loss-arm-relations}
	\end{subequations}
	where we have defined
	\begin{equation}
		\mathbb{P}(\delta)=\mathbb{O}(\delta)e^{i\Omega\tau_{\text{arm}}}
		\label{eq:loss-P}
	\end{equation}
	and $R_{e},T_{e}$ are the amplitude reflectivity and transmissivity of the end test mirror, $\tau_{\text{arm}}=L_{\text{arm}}/c$ is the arm cavity global delay, $k_{p}$ is the wave vector of the light field, $E$ is the classical amplitude of the light field and $\hat{x}(\Omega)$ is a small mirror displacement due to the gravitational wave signal. We make the assumption that the gravitational wave signal appears only in the equations for the phase quadrature:
	\begin{equation}
		E=\sqrt{2}E_{\text{ampl}}\begin{pmatrix}
			1 \\
			0
		\end{pmatrix},
		\label{eq:loss-E}
	\end{equation}
	where the amplitude $E_{\text{ampl}}$ which is connected to the arm cavity light power
	\begin{equation}
		P_{\text{arm}}=\dfrac{\hbar k_{p}c}{2}\lvert E_{\text{ampl}}\rvert^{2}.
		\label{eq:loss-Parm}
	\end{equation}
	We repeat the last step and insert $\hat{d}$ from Eq. \ref{eq:loss-tr-sq-input-output-b} into the Eqs. \ref{eq:loss-arm-relations} to obtain the input-output relations for the entire system:
	\begin{equation}
		\hat{b}=-\mathbb{R}\hat{a}+\mathbb{T}\hat{v}+\mathbb{Z}\hat{x}(\Omega)+\mathcal{L}_{1}\hat{n}_{1}+\mathcal{L}_{2}\hat{n}_{2},
		\label{eq:loss-bsol}
	\end{equation}
	where we have introduced the transfer functions
	\begin{subequations}
		\begin{align}
			&\mathbb{R}=\grave{\mathbb{R}}_{b}-R_{e}\grave{\mathbb{T}}_{b}\mathbb{L}_{c}\mathbb{P}^{2}(\delta)\grave{\mathbb{T}}_{d},\\
			&\mathbb{T}=T_{e}\grave{\mathbb{T}}_{b}\mathbb{L}_{c}\mathbb{P}(\delta),\\
			&\mathbb{Z}=2k_{p}R_{e}\grave{\mathbb{T}}_{b}\mathbb{L}_{c}\mathbb{P}(\delta)\mathbb{O}(\pi/2)E,\\
			&\mathcal{L}_{1}=\grave{\mathcal{L}}_{b1}+R_{e}\grave{\mathbb{T}}_{b}\mathbb{L}_{c}\mathbb{P}^{2}(\delta)\grave{\mathcal{L}}_{d1},\\
			&\mathcal{L}_{2}=\grave{\mathcal{L}}_{b2}+R_{e}\grave{\mathbb{T}}_{b}\mathbb{L}_{c}\mathbb{P}^{2}(\delta)\grave{\mathcal{L}}_{d2},\\
			&\mathbb{L}_{c}=\left(\mathbb{I}-R_{e}\mathbb{P}^{2}(\delta)\grave{\mathbb{R}}_{d}\right)^{-1}.
		\end{align}
		\label{eq:loss-b-TF}
	\end{subequations}
	
	\subsubsection{\label{app:subsubsec:radiation-pressure}Radiation pressure}
	Radiation pressure is the effect that electromagnetic radiation applies a force to an object from which it is reflected \cite{Nichols1903}. Radiation pressure causes a constant displacement of a test mirror, which can be compensated with an active feedback control, and a random displacement due to the uncertainty in the amplitude quadrature of the light. This uncertainty leads to the quantum radiation pressure noise.\\
	Light, which is reflected by a perfectly reflective mirror, exerts the radiation pressure force $F_{\text{rp}}$ on this mirror \cite{Nichols1903}:
	\begin{equation}
		F_{\text{rp}}=\dfrac{2P}{c},
		\label{eq:loss-Frp}
	\end{equation}
	where $P$ is the power of the light. From Eq. \ref{eq:loss-Frp}, we can see that the impact of the radiation pressure force only plays a role for the input and end test mirror. This is because we have a high light power inside the arm cavities and therefore a high radiation pressure force acting on the mirrors of the arm cavities. The impact of the radiation pressure force on the signal extraction mirror can be neglected since we have a low light power inside the signal extraction cavity. We assume the radiation pressure force acting on the input and end test mirror to be equal. This allows us to assume the input test mirror to be fixed and instead twice the back action applied on the end test mirror \cite{Buonanno2003}. This approximation holds as long as the amplitudes of the fields acting on the input test mirror and end test mirror are almost equal, which is the case in the single-mode-approximation. With this approximation, we can calculate the radiation pressure force (also called back-action force):
	\begin{equation}
		F_{ba}=\hbar k_{p}\left(E^{\dagger}\hat{e}(\Omega)+R_{e}E^{\dagger}\hat{f}(\Omega)\right).
		\label{eq:loss-Fba}
	\end{equation}
	We can split the back-action force into a noise part $F_{fl}(\Omega)$ and an optical spring force, which depends on the mirror displacement $\hat{x}(\Omega)$ and a spring constant $\kappa(\Omega)$. With the solutions from the previous Section \ref{app:subsubsec:loss}, we obtain the following equations:
	\begin{subequations}
		\begin{align}
			&F_{ba}=F_{fl}(\Omega)-\kappa(\Omega)\hat{x}(\Omega),\\
			&F_{fl}(\Omega)=\hbar k_{p}(1+R_{e}^{2})E^{\dagger}\mathbb{B}\mathbb{P}(\delta)\left(\grave{\mathbb{T}}_{d}\hat{a}+\grave{\mathcal{L}}_{d1}\hat{n}_{1}+\grave{\mathcal{L}}_{d2}\hat{n}_{2}\right)+\hbar k_{p}T_{e}E^{\dagger}\mathcal{L}_{v}\hat{v},\\
			&\mathbb{B}=\mathbb{I}+R_{e}\mathbb{P}(\delta)\grave{\mathbb{R}}_{d}\mathbb{L}_{c}\mathbb{P}(\delta),\\
			&\mathcal{L}_{v}=R_{e}\mathbb{I}+(1+R_{e}^{2})\mathbb{P}(\delta)\grave{\mathbb{R}}_{d}\mathcal{L}_{c}\mathbb{P}(\delta),\\
			&\kappa(\Omega)=-2\hbar k_{p}^{2}R_{e}E^{\dagger}\Bigl(R_{e}\mathbb{O}(\pi/2)E+(1+R_{e}^{2})\mathbb{P}(\delta)\grave{\mathbb{R}}_{d}\mathcal{L}_{c}\mathbb{P}(\delta)\mathbb{O}(\pi/2)E\Bigr).
		\end{align}
		\label{eq:loss-Fba_sol}
	\end{subequations}
	This leads to the following equation of motion for the end test mirror:
	\begin{equation}
		\hat{x}(\Omega)=\chi_{\text{eff}}(\Omega)F_{fl}(\Omega),
		\label{eq:loss-x}
	\end{equation}
	where we have defined an effective susceptibility 
	\begin{equation}
		\chi_{\text{eff}}(\Omega)=\left(\chi^{-1}(\Omega)+\kappa(\Omega)\right)^{-1}
		\label{eq:loss-chi-eff}
	\end{equation}
	with $\chi(\Omega)=(-m\Omega^{2})^{-1}$ being the mechanical susceptibility and $m$ being the mass of the end test mirror.
	
	\subsubsection{\label{app:subsubsec:detection}Detection}
	The sensitivity of a gravitational wave observatory is reduced by detection loss. Similar to the intra-cavity loss, the detection loss is modelled by a beamsplitter with amplitude transmissivity $\sqrt{\eta}$ and reflectivity $\sqrt{1-\eta}$:
	\begin{equation}
		\hat{\tilde{b}}(\Omega)=\sqrt{\eta}\hat{b}(\Omega)+\sqrt{1-\eta}\hat{n}.
		\label{eq:loss-b-det}
	\end{equation}
	The output $\hat{\tilde{b}}$ is measured by a balanced homodyne detector at homodyne angle $\zeta$. We obtain the following values:
	\begin{flalign}
		\hat{y}(\Omega)=\mathcal{H}^{\text{T}}\hat{\tilde{b}}(\Omega)&=\dfrac{\mathcal{H}^{\text{T}}\left(-\mathbb{R}\hat{a}+\mathbb{T}\hat{v}+\mathcal{L}_{1}\hat{n}_{1}+\mathcal{L}_{2}\hat{n}_{2}\right)}{\mathcal{H}^{\text{T}}\mathbb{Z}}+\dfrac{\sqrt{1-\eta}\mathcal{H}^{\text{T}}}{\sqrt{\eta}\mathcal{H}^{\text{T}}\mathbb{Z}}\hat{n}+\hat{x}(\Omega)
		\label{eq:loss-output-homodyne}
	\end{flalign}
	which we normalized to the mirror displacement and where $\mathcal{H}$ is the homodyne detection transfer vector
	\begin{equation}
		\mathcal{H}=
		\begin{pmatrix}
			\cos(\zeta) \\
			\sin(\zeta)
		\end{pmatrix}.
		\label{eq:loss-homodyne-TV}
	\end{equation}
	The sensitivity can be improved by the injection of squeezed light. In this case, the input field $\hat{a}$ reads as follows:
	\begin{equation}
		\hat{a}=\mathbb{S}_{ext}(\phi_{ext})\hat{a}_{vac},
		\label{eq:loss-a}
	\end{equation}
	where we have introduced the external squeezing matrix
	\begin{equation}
		\mathbb{S}_{ext}(\phi_{ext})=\mathbb{O}(\phi_{ext})
		\begin{pmatrix}
			e^{q_{ext}} & 0 \\
			0 & e^{-q_{ext}}
		\end{pmatrix}
		\mathbb{O}(-\phi_{ext})
		\label{eq:loss-sq-ext}
	\end{equation}
	and $\hat{a}_{vac}$ is the vaccum field before squeezing, $q_{ext}$ is the external squeezing factor and $\phi_{ext}$ is the external squeezing angle.
	The remaining fields $\hat{v}$, $\hat{n}$, $\hat{n}_{1}$, and $\hat{n}_{2}$ are in the vacuum state.\\
	We get the power spectral density:
	\begin{flalign}
		S_{x}(\Omega)&=S_{xx}(\Omega)+2\,\text{Re}\left[\chi_{\text{eff}}^{*}(\Omega)S_{xF}(\Omega)\right]+\lvert\chi_{\text{eff}}(\Omega)\rvert^{2}S_{FF}(\Omega),
		\label{eq:loss-psd-x}
	\end{flalign}
	where we have defined
	\begin{flalign}
		S_{xx}(\Omega)&=\dfrac{\mathcal{H}^{\text{T}}\left(\mathbb{R}\mathbb{S}_{ext}\mathbb{S}^{\dagger}_{ext}\mathbb{R}^{\dagger}+\mathbb{T}\mathbb{T}^{\dagger}+\mathcal{L}_{1}\mathcal{L}^{\dagger}_{1}+\mathcal{L}_{2}\mathcal{L}^{\dagger}_{2}\right)\mathcal{H}}{\lvert \mathcal{H}^{\text{T}}\mathbb{Z}\rvert^{2}}+\dfrac{1-\eta}{\eta \lvert \mathcal{H}^{\text{T}}\mathbb{Z}\rvert^{2}},
		\label{eq:loss-psd-xx}
	\end{flalign}
	\begin{flalign}
		S_{xF}(\Omega)&=\dfrac{\hbar k_{p}}{\mathcal{H}^{\text{T}}\mathbb{Z}}\Big[(1+R_{e})^{2}\mathcal{H}^{\text{T}}\Bigl(-\mathbb{R}\mathbb{S}_{ext}\mathbb{S}^{\dagger}_{ext}\grave{\mathbb{T}}^{\dagger}_{d}+\mathcal{L}_{1}\grave{\mathcal{L}}^{\dagger}_{d1}+\mathcal{L}_{2}\grave{\mathcal{L}}^{\dagger}_{d2}\Bigr)\mathbb{P}^{\dagger}(\delta)\mathbb{B}^{\dagger}E+T_{e}\mathcal{H}^{\text{T}}\mathbb{T}\mathcal{L}^{\dagger}_{v}E\Big],
		\label{eq:loss-psd-xF}
	\end{flalign}
	and
	\begin{flalign}
		S_{FF}(\Omega)=\hbar^{2}k_{p}^{2}(1+R_{e}^{2})^{2}E^{\dagger}\mathbb{B}\mathbb{P}(\delta)\Bigl(\grave{\mathbb{T}}_{d}\mathbb{S}_{ext}\mathbb{S}^{\dagger}_{ext}\grave{\mathbb{T}}^{\dagger}_{d}+\grave{\mathcal{L}}_{d1}\grave{\mathcal{L}}^{\dagger}_{d1}+\grave{\mathcal{L}}_{d2}\grave{\mathcal{L}}^{\dagger}_{d2}\Bigr)\mathbb{P}^{\dagger}(\delta)\mathbb{B}^{\dagger}E\notag\\+\hbar^{2}k_{p}^{2}T_{e}^{2}E^{\dagger}\mathcal{L}_{v}\mathcal{L}^{\dagger}_{v}E.
		\label{eq:loss-psd-x-FF}
	\end{flalign}
	From Eq. \ref{eq:loss-psd-x}, we can compute the power spectral density normalized to the gravitational wave strain yielding, where we take into account the effects of high-frequency corrections \cite{Rakhmanov2008}:
	\begin{equation}
		S_{h}(\Omega)=S_{x}(\Omega)\dfrac{4}{m^{2}L_{\text{arm}}^{2}\Omega^{4}\lvert\chi_{\text{eff}}(\Omega)\rvert^{2}}\dfrac{\Omega^{2}\tau_{\text{arm}}^{2}}{\sin^{2}(\Omega\tau_{\text{arm}})}.
		\label{eq:loss-psd-h}
	\end{equation}		

	\bibliography{ManualSupp, TRQE_paper, JoeBibtex}

\end{document}